# A bootstrap analysis for finite populations

Tina Nane [1] and Kasper Kooijman[2]

[1] g.f.nane@tudelft.nl
Delft University of Technology, Delft (The Netherlands)

[2] kasper_kooijman@live.nl
Delft University of Technology, Delft (The Netherlands)

**Abstract**
Bootstrap methods are increasingly accepted as one of the common approaches in constructing confidence intervals in bibliometric studies. Typical bootstrap methods assume that the statistical population is infinite. When the statistical population is finite, a correction needs to be applied in computing the estimated variance of the estimators and thus constructing confidence intervals. We investigate the effect of overlooking the finiteness assumption of the statistical population using a dataset containing all articles in Web of Science (WoS) for Delft University of Technology from 2006 until 2009. We regard the data as our finite statistical population and consider simple random samples of various sizes. Standard bootstrap methods are firstly employed in accounting for the variability of the estimates, as well as constructing the confidence intervals. The results unveil two issues, namely that the variability in the estimates does not decrease to zero as the sample size approaches the population size and that the confidence intervals are not valid. Both issues are addressed when accounting for a finite population correction in the bootstrap methods.

**Introduction**

The bootstrap (Efron, 1979) has become one of the most well-known statistical methods, despite the futility implied by the 'bootstrap' action, that is, of lifting oneself by using the bootstraps only. Nonetheless, the expressive terminology has been chosen to suggest a process done without external help. Starting from the data at hand, which is regarded as a sample from a statistical population, one is able, by resampling the data, to provide confidence intervals and means of measuring the accuracy of the sample estimates. Since the bootstrap was firstly proposed and studied, constant effort was put into improving the techniques, relaxing the assumptions and accounting for the particularities of the data at hand. This, together with the enhanced computational capabilities lead to the current popularity of the bootstrap.

Bootstrap are increasingly accepted as one of the common approaches to construct confidence intervals in various science fields, including citation analysis and bibliometric studies. Chen, Jen & Wu (2014), Rahman et al. (2016), Fairclough & Thelwall (2015), Williams & Bornmann (2016), Bornmann & Daniel (2007), Waltman & Costas (2013), Costas, Zahedi & Wouters (2014) are just a few recent publications that have included bootstrap methods in their analysis. Bootstrap methods are also used in accounting for the variability in the research output in the Leiden Ranking (Waltman et al., 2012).

The standard bootstrap methods, as well as various other standard statistical tools rely on the implicit assumption that the data at hand is a sample from an infinite population. "The idea of infinite hypothetical population is, I believe, implicit in all statements involving mathematical probability" (Fisher, 1925), and therefore in statistics. This is however rarely mentioned in practice by statisticians. This of course has to do with the purpose of the statistical analysis and the nature of the target population, as emphasized in Nane (2016) as a comment to Williams & Bornmann (2016).

Though implicit, numerous statistical populations in citation analysis are assumed to be infinite, albeit data on publications or researchers. The target population can also be regarded as a underlying stochastic process, whose realizations are observable and constitute a statistical population, see, for example Claveau (2016). The concept of super-population emerges therefore naturally.

Finite target populations arise in numerous applications, for example in survey analysis, biology, management, etc. When the target population is finite, the statistical methods need to be adapted. Hajek (1960, 1981) has initiated the work on the finite population statistics and provided important asymptotic results, as well as advanced the idea of sampling within finite populations. This study follows the discussion on the paper of Williams & Bornmann (2016). In the comment, Nane (2016) drew attention on the particularity of the data at hand, that is the finite population and the necessary correction implied by this assumption.

The purpose of this study is twofold. First of all, several techniques of performing bootstrap are considered. Results are investigated and conclusions are drawn based on the variability of the estimates and confidence intervals with respect to length and accuracy. Secondly, we will investigate the effect of not taking into account the finiteness of the population and we will, in turn, account for the finiteness of the population at hand.

The study relies on the Delft University of Technology (TU Delft) data, that is all the articles from Web of Science (WoS) for which at least one author is affiliated to TU Delft, as collected for Leiden Ranking 2011/2012. We will perform a simulation study where the data at hand will act as the statistical, finite, population and various samples, using a simple random sampling without replacement will be drawn from this population. The 6224 articles in our data have been published between 2006 and 2009 and their citations have been counted until the end of 2010. The citation performance of the articles are evaluated using two well-known indicators, MNCS and PP(top 10%). MNCS is the average of the field-normalized citation score of each publication. PP(top 10%) accounts for the proportion of publications that are in the top 10% most cited in their field. More details about the indicators can be found in Waltman et al. (2012). The population MNCS is 1.275, that is the average of the field-normalized citation scores of the TU Delft publications is 1.275, which is higher than the world average and the PP(top 10%) is 13.7%, which asserts that 13.7% of TU Delft's publications are in the top 10% most cited publications in their field. These values will be regarded as the true parameter values and compared with estimates and confidence intervals resulting from various bootstrap methods.

The paper is organized in three subsequent sections. The following section describes the methods used for our analysis. It includes and compares standard bootstrap methods, as well as the bootstrap methods which account for the finite population. The main results of the analysis are presented in the subsequent section. Finally, the conclusions are drawn and the discussion is carried.

## Methods

We investigate the effects of the finite population setting on the estimation of two well-known citation indicators, MNCS and PP(top 10%), as well as on the construction of confidence intervals. Standard bootstrap methods will be reviewed and compared in the following subsection. The adjustments entailed by the finite population setting will be considered and integrated in bootstrap methods.

*Bootstrap methods*

Bootstrap has become a standard statistical method for providing a measure of variation for sample estimates and computing confidence intervals. Since it was first advanced by Efron (1979), numerous bootstrap methods have been proposed.

Given an independent and identically distributed sample $s = \{x_1, x_2, \ldots, x_n\}$ from an unknown distribution $F$, we are interested in estimating the parameter $\theta$ of the distribution $F$, typically the mean. The estimate of $\theta$ is denoted by $\hat{\theta}$, the sample mean estimate of the mean of the distribution. $B$ samples, denoted $s_1^*, s_2^*, \ldots, s_B^*$, are drawn, with replacement from the initial sample $s$. These samples represent the bootstrap samples and the size of each bootstrap sample is equal to $n$. A bootstrap parameter $\theta_i^*$ is computed for every bootstrap sample $s_i^*$, for $i = 1, \ldots, n$.

The variance of the estimator $\hat{\theta}$ is estimated by the bootstrap variance estimator. In turn, the bootstrap variance estimator has been shown to converge asymptotically to the sample variance of the bootstrap estimator.

This bootstrap methods is also referred to as the nonparametric bootstrap, since no parametric distribution is assumed for $F$. Assuming a certain parametric family for $F$ and drawing the bootstrap samples from a parametric distribution where the parameters have been estimated from the sample $s$ are often referred to as the parametric bootstrap. We will not include the parametric bootstrap in our analysis.

The standard approach for computing confidence intervals is based on the central limit theorem, and uses the sample variance bootstrap estimator and quantiles of the standard normal distribution. These confidence intervals will be called asymptotic confidence intervals. Confidence intervals can also be constructed based on the percentiles of the bootstrap distribution, suggestively named percentile confidence intervals. The percentiles confidence intervals can be quite inaccurate in the case of skewed distributions. A bias corrected version (BCa) of the bootstrap has been proposed to compute confidence intervals (Efron, 1984). DiCiccio & Efron (1996), for example, provide a detailed theoretical analysis of the accuracy of these bootstrap confidence intervals.

Though implicit, standard bootstrap methods assume that the statistical population from which the data at hand was sampled is infinite. In the following section, we will see that, in practice, bootstrap methods fail to provide adequate results when this assumption is violated.

*Finite population correction*

The finite population correction assumes that the population size is known. Let $N$ be the population size and let $n$ be the sample size. Then the sampling fraction is defined as $f = n/N$ and the finite population correction is given by $1 - f$, see, for example, Davidson and Hinkley (1997). If the variance of the bootstrap estimator is denoted by $V^*$, then the finite population correction of the variance estimator is $(1 - f)V^*$, from where it becomes obvious that it converges to zero, as $n \to N$. A straightforward bias-adjusted variance estimator is given by

$$V^{*\prime} = \frac{N - n}{N - 1} V^*.$$

When estimating means, the standard error of the mean for finite populations is given

$$\frac{\sigma}{\sqrt{n}}\sqrt{\frac{N-n}{N-1}}.$$

The second term of the product is often referred to as the finite population correction factor (*fpc*) and is also used in deriving the standard error of the proportion for finite populations. Finally, it needs to be emphasized that the finite population correction depends on the sampling design. As mentioned beforehand, the employed sampling scheme here is simple random sampling without replacement. If a different sampling design is chosen or has been already employed when collecting the data, then the finite population correction needs to be adjusted accordingly.

The question is of course when does the finite population make a significant difference. Davidson and Hinkley (1997) advise to use the finite population correction when $f \geq 0.1$. That is to say, if $f < 0.1$, then $n$ is relatively small compared to $N$ and the correction factor can be ignored.

*Bootstrap methods for finite populations*

Several methods have been proposed in the literature for the bootstrap while taking into account the finiteness of the population. Mashreghi, Haziza & Léger (2016) have recently provided a survey of bootstrap methods for finite populations in the context of survey data and for various sampling techniques. We will investigate two bootstrap methods, namely the pseudo-population bootstrap and the direct bootstrap for simple random sampling without replacement.

The pseudo-population bootstrap method creates, by resampling, a pseudo-population with the same size as the actual finite population. Bootstrap samples, equally sized as the original sample are subsequently drawn from the resulting pseudo-population. By employing the same sampling design in drawing the bootstrap samples from the pseudo-population as for drawing the initial sample, the method ensures that the bootstrap variance estimator encompasses the finite population correction factors. The pseudo-population method has been first proposed by Gross (1980) and several adaptations have been developed ever since. We will employ in our analysis the method proposed by Booth, Butler & Hall (1994).

For comparison reasons, we have also employed the direct bootstrap method for finite populations. Direct bootstrap methods are grouped in a category that does not employ the creation of the pseudo-population but mimic the idea and adjust the original proposal of Efron (1979), where bootstrap samples are directly drawn from the data at hand. The adjustments are made "so that the bootstrap variability reflects the sampling variability of the original sample design" (Mashreghi, Haziza & Léger, 2016) while accounting for the finite population setting. In the direct method proposed by Sitter (1992), resamples of smaller size are taken from the initial sample without replacement for a number of times that depends on the size of the original sample and the resample and the finite population correction factor. The bootstrap sample is constructed by concatenating all these resamples. The motivation and technical details, along with two other direct methods are extensively studied in Mashreghi, Haziza & Léger (2016). A comparison between the direct bootstrap methods and the standard bootstrap proposed by Efron (1979) shows how Efron's bootstrap method, in the absence of a finite population correction factor, leads to an overestimate of the variance.

Confidence intervals can be constructed using the same techniques as when using the standard bootstrap methods. Parametric confidence intervals will be constructed, as well as bootstrap-t confidence intervals. The bootstrap-t confidence intervals use quantiles of the bootstrap distribution $(\hat{\theta} - \theta)/\sqrt{\hat{V}}$, where $\hat{V}$ is an estimate of the variance of the estimator $\hat{\theta}$ instead of quantiles of the standard normal distribution. Moreover, we will employ standard percentile confidence intervals that make use of the bootstrap distribution of $\hat{\theta}^*$. As noted by Mashreghi, Haziza and Léger (2016), it is worth mentioning that, in practice, the percentile confidence intervals over-cover the true parameter $\theta$. This is generated by the overall high dispersion of the bootstrap estimators $\hat{\theta}^*$ compared to the dispersion of $\hat{\theta}^* - \theta^*$ that is used in computing the bootstrap-t confidence intervals.

*Finite population in statistical software packages*

The assumption of finite population is usually contained within the survey analysis of the statistical software programs. The finite population correction is included in various statistical software programs as Stata, SAS and R. In Stata, it is included as an option in the survey design functions. SAS also has the finite population correction factor contained within the survey function PROC SURVEYMEANS. R has various packages that account for finite population correction, including survey and PracTools. In Stata, the variance estimation and confidence intervals for finite population are obtained using bootstrap procedure proposed by McCarthy
& Snowden (1985) and Rao & Wu (1988) in the class of direct bootstrap methods. The differences between the two methods and the method of Sitter (1992) employed in our analysis are quite technical and can be found in the survey of Mashreghi, Haziza & Léger (2016). To the best of our knowledge there are no bootstrap methods for finite populations implemented in SAS. In R, the survey package also the direct bootstrap method of Rao & Wu (1988). To the best of our knowledge, the pseudo-population bootstrap and the direct bootstrap method that we used in our analysis have not been implemented yet in a package in R or elsewhere.

**Results**

This section includes the analysis of the standard bootstrap method for various sample sizes, as well as two bootstrap methods that allow for the finite population assumption.

*Standard bootstrap methods*

To study the sample estimates and to compare the different 95% confidence intervals, a sample of size $n = 100$ has been drawn from our statistical population. The table below shows the results for the two indicators, MNCS and PP(top 10%).

Table 1. MNCS and PP(top 10%) estimates based on a sample n=100 and B=1000 bootstrap samples, and 95% confidence intervals using an asymptotic, percentile and BCa method.

|  | *Estimate* | *Variance of Estimator* | *Asymptotic CI* | *Percentile CI* | *BCa CI* |
|---|---|---|---|---|---|
| MNCS | 1.22 | 0.21 | (0.88;1.71) | (0.91;1.74) | (0.96;1.84) |
| PP(top 10%) | 12.58% | 0.03 | (6.44%;18.89%) | (6.69%;19.12%) | (7.49%;20.20%) |

The sample MNCS is 1.22 and the estimate for PP(top 10%) is 12.58%, which are relatively different from the true values of MNCS is 1.275 and PP(top 10%) is 13.65%. The variance of the two estimators is estimated using the bootstrap method. As expected, the variance for the MNCS is significantly higher than the variance for the PP(top 10%). This difference reflects the robustness of the PP(top 10%) as compared to the MNCS. The three confidence intervals are rather wide, showing the high uncertainty around the estimates. Finally, it is notable that all confidence intervals contain the true values.

Obviously these results depend on the particular initial sample, so an analysis of the bootstrap has been performed by repeating the bootstrap procedure for a large number of times. Specifically, for 1000 times, a sample of size 100 has been drawn. The MNCS and PP(top 10%) estimate has been computed, along with the variance of the estimators and the b o o t s t r a p confidence intervals for the 1000 samples. The results are depicted in Table 2.

**Table 2. Average length and coverage of the true value for asymptotic, percentile and BCa confidence intervals (CI) for MNCS and PP(top 10%) based on a sample n=1000 and B=1000 bootstrap samples.**

|  | Variance of Estimator | Type CI | Average length | Coverage |
|---|---|---|---|---|
| MNCS | 0.17 | Asymptotic | 0.71 | 89.8% |
|  |  | Percentile | 0.76 | 90.7% |
|  |  | BCa | 0.81 | 91.9% |
| PP(top 10%) | 0.03 | Asymptotic | 12.89 | 91.5% |
|  |  | Percentile | 12.9 | 93.7% |
|  |  | BCa | 13.27 | 95.8% |

The average of the estimated variance of the estimator has been obtained by averaging the estimated variances of the two estimators over the bootstrap samples. We note small differences from the estimated variance of the estimator using one sample in Table 1. The average length provides the average over the length of the confidence intervals obtained from applying the three bootstrap methods. Furthermore, we computed the coverage probability of the confidence intervals. That is, we checked whether the true MNCS and PP(top 10%) values were contained in each resulting confidence interval. We concluded that, for example, the true MNCS values was contained in 907 out of 1000 percentile confidence intervals. Similarly, 958 times was PP(top 10%) true value contained in the BCa confidence intervals.

Since we computed 95% confidence intervals, we expect a 95% coverage of the confidence intervals. The results in Table 2 show that the 95% confidence intervals are mostly under- covering the true values. This effect is highly likely to be due to the small sample size. We will hence investigate the coverage probability, as well as the average length and average of the estimated variance of the estimator for larger samples, namely for $n = 1000$.

**Table 3. Average length and coverage of the true value for asymptotic, percentile and BCa confidence intervals (CI) for MNCS and PP(top 10%) based on a sample n=1000 and B=1000 bootstrap samples.**

|  | Average of Variance of Estimator | Type CI | Average length | Coverage |
|---|---|---|---|---|
| MNCS | 0.07 | Asymptotic | 0.27 | 96.4% |
|  |  | Percentile | 0.27 | 96.1% |
|  |  | BCa | 0.28 | 96.8% |

| PP(top 10%) | 0.01 | Asymptotic | 4.13 | 97.9% |
|---|---|---|---|---|
| | | Percentile | 4.10 | 98.2% |
| | | BCa | 4.13 | 99.1% |

The results in Table 3 show a rapid decrease of the variance of estimators and average length of the confidence intervals as the sample size increases. The average of the estimated variance of the estimators has decreased to less than a half. The average length of the confidence intervals has also decreased considerably. Finally, we notice an increase in the confidence intervals coverage. The high coverage values, which might be interpreted as a very good performance, show that the confidence intervals are conservative and over-cover the true parameter.

It also raises a question on the validity if the confidence interval. A confidence interval is said to be valid if the coverage of the interval converges to the true coverage ( of 95% in our case) for a very large number of repetitions of the procedure. When repeating the procedure for a number of times, the coverage probabilities have slightly changed.

Furthermore, when repeating the bootstrap methods for increasing sample sizes, the variance of the estimators and the length of the confidence intervals decreases. Since the asymptotic confidence intervals are derived using the variance of the estimator, we will investigate the influence of the sample size on the length of the asymptotic confidence intervals. Figure 1 below depicts the length of the confidence intervals for sample sizes until 4000 observations and for a sample size equal the statistical population of 6224 observations.

**Figure 1. Length of asymptotic confidence intervals for MNCS for different sample sizes using the standard bootstrap method (left) and length of 100 MNCS confidence intervals using a sample size equal to the population size (6224).**

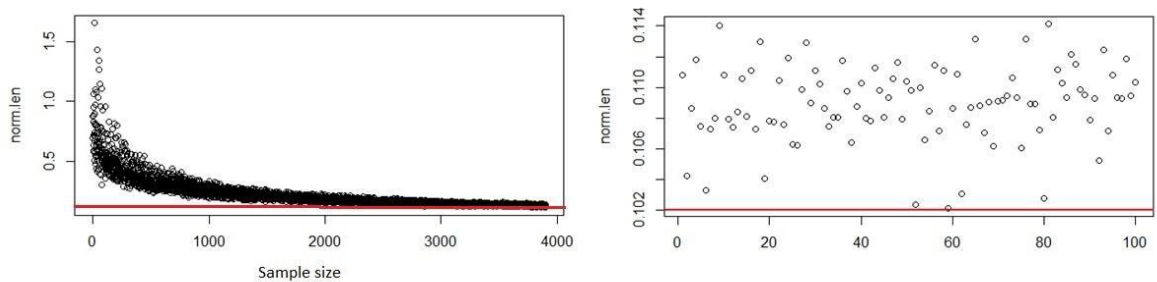

The two plots show that, although decreasing, the variance of the MNCS estimator and hence the confidence intervals do not decrease to zero, even when the sample size equals the size of the statistical population, as depicted in the right plot. When the sample is, in fact, the statistical population, the estimate is as a matter of fact the true parameter and has zero variance. Nonetheless, this is not reflected by the right plot of Figure 1. The reason why the variance and hence the length of the confidence intervals is not decreasing to zero is because the sample is assumed to be drawn from an infinite population. A finite population correction therefore needs to be applied. The bootstrap methods that account for finite populations are employed in the following subsection.

*Bootstrap methods for finite population*

We will use the pseudo-population and the direct method described in the previous section for constructing asymptotic, bootstrap-t and percentile confidence intervals for the MNCS and PP(top 10%) estimators. The results for a simple random sample of size 500 are included in Table 4.

**Table 4. Coverage probability and average length for the asymptotic, bootstrap-t and percentile confidence intervals using the pseudo-population bootstrap (PPB) and the direct bootstrap.**

| *Table* | *Bootstrap* | *Type CI* | *Average length* | *Coverage* |
|---|---|---|---|---|
| MNCS | PPB | Asymptotic | 0.37 | 93.1% |
|  |  | Bootstrap-t | 0.36 | 92.4% |
|  |  | Percentile | 0.36 | 94.9% |
|  | DB | Asymptotic | 0.37 | 95.2% |
|  |  | Bootstrap-t | 037 | 92.3% |
|  |  | Percentile | 0.37 | 95.1% |
| PP(top 10%) | PPB | Asymptotic | 5.58 | 95.6% |
|  |  | Bootstrap-t | 5.85 | 95.2% |
| PP(top 10%) |  | Percentile | 5.57 | 95.8% |
|  | DB | Asymptotic | 5.67 | 94.4% |
|  |  | Bootstrap-t | 5.68 | 94.1% |
|  |  | Percentile | 5.67 | 95.1% |

We observe similar results in terms of the average length of the intervals for the two bootstrap methods. The coverage of the confidence intervals suggest, in general, a better coverage for the PP(top 10%) than for MNCS. When using the pseudo-population bootstrap, the MNCS seems to be under-covered by the confidence intervals, where the highest coverage probability is 94.9% for the percentile confidence intervals. Using the direct method produces confidence intervals that cover the true MNCS value close to the nominal coverage. For PP(top 10%), the coverage probabilities for the pseudo-population confidence intervals are closer to the nominal value. The closest coverage probabilities to the nominal value, for both MNCS and PP(top 10%), and for both bootstrap methods are obtained by the percentile confidence intervals.

**Table 5. Coverage probability and average length for the asymptotic, bootstrap-t and percentile confidence intervals using the pseudo-population bootstrap (PPB) and the direct bootstrap.**

| *Table* | *Bootstrap* | *Type CI* | *Average length* | *Coverage* |
|---|---|---|---|---|
| MNCS | PPB | Asymptotic | 0.24 | 93.1% |
|  |  | Bootstrap-t | 0.24 | 92.4% |
|  |  | Percentile | 0.24 | 94.9% |
|  | DB | Asymptotic | 0.25 | 95.2% |
|  |  | Bootstrap-t | 024 | 92.3% |
|  |  | Percentile | 0.25 | 95.1% |
| PP(top 10%) | PPB | Asymptotic | 2.78 | 95.6% |
| | | Bootstrap-t | 2.73 | 95.2% |
| | | Percentile | 2.85 | 95.8% |
| | DB | Asymptotic | 2.86 | 94.4% |
| | | Bootstrap-t | 2.86 | 94.1% |
| | | Percentile | 2.85 | 95.2% |

Table 5 shows the analysis for a simple random sample of 1000 observations. The average length is decreasing when comparing with Table 4. Nonetheless, the coverage probability fluctuates around the nominal coverage probability of 95%, sustaining the validity of the confidence intervals produced. Once more, the percentile confidence intervals are the closest to the true coverage probability.

To investigate the change in the length of the confidence intervals with respect to the sample size, we performed simulations for sample sizes varying from 100 to the entire statistical population.

**Figure 2.** Length of asymptotic and percentile confidence intervals (CI), for n=100, 500, 1000, 2000, 4000 and 6224 using the direct bootstrap method (left), pseudo-population bootstrap (middle) and the percentile confidence intervals produced with boot.ci function in R (right).

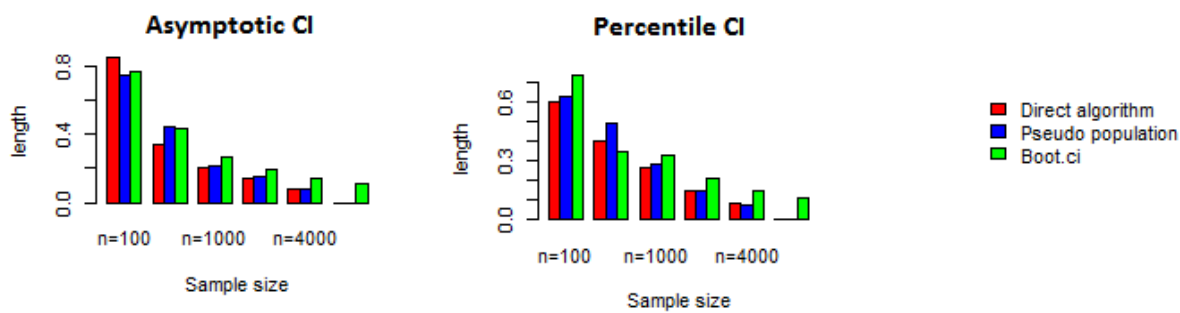

Figure 2 above shows the length of the asymptotic and percentile confidence intervals using samples with size varying from 100 (leftmost) to 6224, the size of the statistical population. For each sample size, the direct method of Sitter (1992), the pseudo-population of Booth, Butler & Hall (1994) and the standard bootstrap method from Davidson & Hinkley (1997) implemented in the statistical R package boot are used to compute confidence intervals. As expected, the length of the confidence intervals decreases as the sample size increases. It is notable however that the standard bootstrap method used by the boot.ci function in R does not decrease to zero.

## Conclusions and discussion

The purpose of this study is to draw attention and provide a detailed analysis for the finite populations when bootstrap techniques are employed. The confidence intervals constructed using standard bootstrap techniques have been shown to over-cover the true parameter as the sample size increases. The over-coverage appears for sample sizes of less than 16% of the statistical population. The over-coverage poses serious questions on the validity of the confidence intervals, which need to have a coverage around the nominal value of 95%. The main reason for this phenomenon is that the variability in the estimator does not take into account the finiteness of the population. The analysis shows that the estimated variance of MNCS and PP(top 10%) estimates decreases as the sample size increases. Despite small, the estimates of the variance of MNCS and PP(top 10%) influence significantly the coverage of the confidence intervals.

Finite populations depend on the choice of target population and can definitely occur in citation analysis. An example at hand is the use of the bootstrap methods to compute the stability intervals in the Leiden Ranking (Waltman et al., 2012). All publications of universities in WoS are regarded as the (finite) target population from which samples of size 1000 are drawn, with replacement, in order to compute the stability intervals. It is notable that the widest stability intervals are encountered for universities with the smallest output, of around 1000 publications. On the other hand, for these universities, the samples are the largest with respect to the population size, which should translate into smaller variation of the indicators and hence smaller stability intervals. A finite population correction for the variability of the indicators, as well as bootstrap methods for finite population would account for the relationship between the sample size and the population size.

The results suggest that the finite population correction needs to be incorporated in any proper analysis. Nonetheless, the scarce and limited availability of bootstrap methods for finite populations make the process difficult. Despite the large application of these techniques, they are contained in packages and functions devoted to survey analysis. Though far from exhaustive, the bootstrap methods presented in this study are theoretically accurate, as presented by Mashreghi, Haziza & Léger (2016). We need to emphasize though that a large number of variations of the bootstrap are available, depending on the focus of the analysis and the data at hand.

First of all, the sampling procedure itself can lead, in the case of finite populations to different bootstrap methods. In our study, we used simple random sampling without replacement. Other methods are available, including mirror-match bootstrap and superpopulation bootstrap. The details of these methods can be found in Sitter (1992) and Davidson & Hinkley (1996). All in all to show that finite population sampling is not a trivial matter and the effect of the sampling design should be accounted for. In practice, researchers can opt for stratified sampling, unequal sampling, etc. The study of the effect of the sampling design on the variability of the bibliometric indicators would be of interest.

An important assumption of the methods presented is that the data at hand are independent and identically distributed. It has been frequently shown that the independence assumption is unrealistic and one can argue that, in this analysis, the citation counts of the publications at TU Delft are not independent. Bootstrap methods are available for dependent, though identically distributed data (Gonçalves & Politis, 2011) and it would be very interesting to investigate the existing dependencies and also the effect of those dependencies on the estimates and confidence intervals.

Last, but definitely not the least, we reinforce the concepts of statistical sample and statistical population. A very important question evidently aims for the target of the analysis and therefore target population. As emphasized in Nane (2016), the target population triggers the analysis and therefore the validates the use of finite population correction and bootstrap methods of computing the confidence intervals.

## Acknowledgments

The authors are grateful to CWTS, Leiden University, for providing the data for this study.

**References**

Booth, J. G., Butler, R. W., & Hall, P. (1994). Bootstrap methods for finite populations. *Journal of the American Statistical Association*, *89*(428), 1282-1289.

Bornmann, L. & Daniel, H.D. (2007). Multiple publication on a single research study: does it pay? The influence of number of research articles on total citation counts in biomedicine. *Journal of the American Society for Information Science and technology*, *58*(8), 1100-1107.

Claveau, F. (2016). There should be no mystery: A comment on sampling issues in bibliometrics. *Journal of Informetrics,* 10, 1233-1240.

Chen, K.-M., Jen, T.-H. & Wu, M. (2014). Estimating the accuracies of journal impact factor through bootstrap. *Journal of Informetrics*. 8, 181-196.

Costas, R., Zahedi, Z. & Wouters, P. (2014). Do "altmetrics" correlate with citations? Extensive comparison of altmetric indicators with citations from a multidisciplinary perspective. *Journal of the Association for Information Science and Technology*. 66(10), 2003-2019.

Davidson, A.C. & Hinkley, D.V. (1997). *Bootstrap Methods and Their Applications*. Cambridge. DiCiccio, T.J. & Efron, B. (1996). Bootstrap confidence intervals. *Statistical Science.* 3, 189-228. Efron, B. (1979). Bootstrap methods: another look at the jackknife. *The Annals of Statistics*, 1-26. Efron, B. (1984). Better bootstrap confidence intervals. *Tech. Rep. Standford Univ. Dept. Stat.*

Fairclough, R. & Thelwall, M. (2015). More precise methods for national research citation impact comparisons. *Journal of Informetrics*. 9, 895-906.

Fisher, R.A. (1925). Theory of statistical estimation. *Mathematical Proceedings of the Cambridge Philosophical Society.* 22(05), 700-725.

Gonçalves, S., & Politis, D. (2011). Discussion: Bootstrap methods for dependent data: A review. *Journal of the Korean Statistical Society*, *40*(4), 383-386.

Gross, S. (1980). Median estimation in sample surveys. In *Proceedings of the Section on Survey Research Methods* (Vol. 1814184). American Statistical Association Ithaca, NY.

Hájek, J. (1960). Limiting distributions in simple random sampling from a finite population. *Publications of the Mathematics Institute of the Hungarian Academy of Science*, *5*(361), 74.

Hájek, J. (1981). *Sampling from a finite population* (Vol. 37). V. Dupac (Ed.). M. Dekker.

Mashreghi, Z., Haziza, D., & Léger, C. (2016). A survey of bootstrap methods in finite population sampling. *Statistics Surveys*, *10*, 1-52.

McCarthy, P. J., & Snowden, C. B. (1985). The bootstrap and finite population sampling. *Vital and health statistics. Series 2, Data evaluation and methods research*, (95), 1-23.

Nane, G.F. (2016). To infer or not to infer? A comment on Williams and Bornmann. *Journal of Informetrics*. 10(4), 1131-1134.

Rahman, A.I.M.J., Guns, R., Leydesdorff, L. & Engels, T.C.E. (2016). Measuring the math between evaluators and evaluees: cognitive distances between panel members and research groups at the journal level. *Scientometrics*. 109, 1639-1663.

Rao, J. N., & Wu, C. F. J. (1988). Resampling inference with complex survey data. *Journal of the american statistical association*, *83*(401), 231-241.

Sitter, R. R. (1992). A resampling procedure for complex survey data. *Journal of the American Statistical Association*, *87*(419), 755-765.

Thewall, M. & Fairclough, R. (2017). The accuracy of confidence intervals for field normalized indicators. *arXiv preprint arXiv:1703.04031*.

Waltman, L., Calero-Medina, C., Kosten, J., Noyons, E., Tijssen, R. J., Eck, N. J., ... & Wouters, P.(2012). The Leiden Ranking 2011/2012: Data collection, indicators, and interpretation. *Journal of the American Society for Information Science and Technology*, *63*(12), 2419-2432.

Waltman, L. & Costas, R. (2013). F1000 recommendations as a potential new data source for research evaluation: a comparison with citations. *Journal of the Association for Information Science and Technology*. 65(3), 433-445.

Williams, R. & Bornmann, L. (2016). Sampling issues in bibliometric analysis. *Journal of Informetrics*. 10(4), 1225-1232.